\documentclass[twocolumn]{aastex63}

\usepackage{amsmath}
\usepackage{color}

\received{\today}
\revised{\today}
\accepted{\today}
\submitjournal{ApJ}

\shorttitle{Measuring the \hi\ gas content with [\cii]}
\shortauthors{Heintz et al.}

\newcommand{\hi}{H\,{\sc i}}
\newcommand{\cii}{C\,{\sc ii}}
\newcommand{\ci}{C\,{\sc i}}

\begin{document}

\title{Measuring the \hi\ content of individual galaxies out to the epoch of reionization with [\cii]}

\correspondingauthor{Kasper E. Heintz}
\email{keheintz@hi.is}

\author[0000-0002-9389-7413]{Kasper~E.~Heintz}
\affil{Centre for Astrophysics and Cosmology, Science Institute, University of Iceland, Dunhagi 5, 107 Reykjav\'ik, Iceland}
\affil{Cosmic Dawn Center (DAWN), Niels Bohr Institute, University of Copenhagen, Jagtvej 128, 2100 Copenhagen \O, Denmark}

\author[0000-0002-4465-8264]{Darach~Watson}
\affil{Cosmic Dawn Center (DAWN), Niels Bohr Institute, University of Copenhagen, Jagtvej 128, 2100 Copenhagen \O, Denmark}

\author[0000-0001-5851-6649]{Pascal~A.~Oesch}
\affil{Cosmic Dawn Center (DAWN), Niels Bohr Institute, University of Copenhagen, Jagtvej 128, 2100 Copenhagen \O, Denmark}
\affil{University of Geneva, Department of Astronomy, Chemin Pegasi 51, 1290 Versoix, Switzerland}

\author[0000-0002-7064-4309]{Desika~Narayanan}
\affil{Cosmic Dawn Center (DAWN), Niels Bohr Institute, University of Copenhagen, Jagtvej 128, 2100 Copenhagen \O, Denmark}
\affil{Department of Astronomy, University of Florida, 211 Bryant Space Sciences Center, Gainesville, FL 32611 USA}
\affil{University of Florida Informatics Institute, 432 Newell Drive, CISE Bldg E251, Gainesville, FL 32611}

\author{Suzanne~C.~Madden}
\affil{AIM, CEA, CNRS, Universit\'e Paris-Saclay, Universit\'e Paris Diderot, Sorbonne Paris Cit\'e, 91191 Gif-sur-Yvette, France}

\begin{abstract}
The \hi\ gas content is a key ingredient in galaxy evolution, the study of which has been limited to moderate cosmological distances for individual galaxies due to the weakness of the hyperfine \hi\ 21-cm transition.
Here we present a new approach that allows us to infer the \hi\ gas mass $M_{\rm HI}$ of individual galaxies up to $z\approx 6$, based on a direct measurement of the [\cii]-to-\hi\ conversion factor in star-forming galaxies at $z\gtrsim 2$ using $\gamma$-ray burst afterglows. By compiling recent [\cii]-158\,$\mu$m emission line measurements we quantify the evolution of the \hi\ content in galaxies through cosmic time. We find that the \hi\ mass starts to exceed the stellar mass $M_\star$ at $z\gtrsim 1$, and increases as a function of redshift. The \hi\ fraction of the total baryonic mass increases from around $20\%$ at $z = 0$ to about $60\%$ at $z\sim 6$. We further uncover a universal relation between the \hi\ gas fraction $M_{\rm HI}/M_\star$ and the gas-phase metallicity, which seems to hold from $z\approx 6$ to $z=0$. The majority of galaxies at $z>2$ are observed to have \hi\ depletion times, $t_{\rm dep,HI} = M_{\rm HI}/{\rm SFR}$, less than $\approx 2$\,Gyr, substantially shorter than for $z\sim 0$ galaxies.
Finally, we use the [\cii]-to-\hi\ conversion factor to determine the cosmic mass density of \hi\ in galaxies, $\rho_{\rm HI}$, at three distinct epochs: $z\approx 0$, $z\approx 2$, and $z\sim 4-6$. These measurements are consistent with previous estimates based on 21-cm \hi\ observations in the local Universe and with damped Lyman-$\alpha$ absorbers (DLAs) at $z\gtrsim 2$, suggesting an overall decrease by a factor of $\approx 5$ in $\rho_{\rm HI}(z)$ from the end of the reionization epoch to the present. 
\end{abstract}


\section{Introduction} \label{sec:intro}

The first epoch of galaxy formation is governed by the infall of pristine gas. These reservoirs of neutral gas are composed mostly of atomic hydrogen (\hi), which then cools and condenses into molecular gas and initiates star-formation. The \hi\ gas content is therefore a key ingredient in galaxy evolution. In the local Universe the hyperfine \hi\ 21-cm transition has been used as the main tracer of this neutral atomic gas \citep{Zwaan05,Hoppmann15,Jones18HI}, but due to the weakness of the line this approach is only feasible out to $z\approx 0.4$ for individual galaxies \citep{Fernandez16}. 

Similarly, cold molecular hydrogen (H$_2$), is difficult to detect in emission due to its lack of a permanent dipole moment. The recent advent of facilities operating at sub-millimeter wavelengths such as the Atacama Large Millimeter/submillimeter Array (ALMA), however, have enabled exploration of the dusty and molecular gas-phases of galaxies through proxies such as carbon monoxide \citep[CO;][]{Tacconi10,Bolatto13}, neutral atomic carbon \citep[\ci;][]{Papadopoulos04,Walter11,Valentino18}, ionized carbon \citep[\cii;][]{Zanella18,Madden20}, and dust \citep{Magdis12,Scoville14,Scoville16}, which have all been employed as effective tracers of the molecular gas \citep[see also][]{Tacconi20,Valentino20,Harrington21}. The weakness of the hyperfine 21\,cm transition of \hi\ and the crucial importance of \hi\ as the fuel for galaxy evolution, similarly demands a powerful alternative that can probe the neutral atomic \hi\ gas content in the most distant galaxies.   

Here we develop a calibration similar to those utilized in high-redshift galaxy surveys targeting the CO or \ci\ transitions to infer the total molecular or H$_2$ gas mass, but instead calibrated to \hi. The [\cii]-158\,$\mu$m fine-structure transition, often the brightest atomic or ionic line emitted from galaxies, is an attractive candidate for such a calibration. This line originates primarily from the cold, neutral gas so finding a way to calibrate it to trace the overall \hi\ gas content is an important goal. While the line has previously been used as a proxy for the molecular gas content of galaxies \citep{Zanella18,Madden20}, observations \citep{Pineda14,Croxall17,DiazSantos17,Cormier19,Tarantino21} and simulations \citep{Franeck18,Olsen21,RamosPadilla21} show that typically the molecular phases make only a minor contribution to the total [\cii] line luminosity. The association of the [\cii]-158\,$\mu$m line with the diffuse ionized or neutral atomic gas is further supported by the extent of its emission being greater than the star-forming disk, the CO component, and the dust continuum in both low- and high-redshift galaxies \citep{Madden97,Carniani18,Fujimoto19,Harikane20,HerreraCamus21}. A significant fraction of the [\cii] emission is therefore proposed to originate from the extended interstellar gas disk of galaxies.
Determining the [\cii]-to-\hi\ conversion factor thus provides a unique approach to probing the \hi\ gas content in individual galaxies. 

The approach presented here does not rely on any scaling relations or assumptions about the physical state of the gas. The [\cii]-to-\hi\ calibration is derived purely from direct measurements of the relative abundances in the line-of-sight through the interstellar medium (ISM) of galaxies hosting $\gamma$-ray bursts (GRBs). This approach does therefore not necessarily require the [\cii]-emitting gas to be physically associated with or predominantly originate in the \hi\ gas reservoirs, but only provides an average abundance ratio of the two throughout the ISM. We present the GRB sample and outline the methodology used to derive the [\cii]-to-\hi\ conversion factor in Section~\ref{sec:data}. In Section~\ref{sec:res} we describe the sample compilation of [\cii]-emitting galaxies at $z\gtrsim 2$ and quantify the evolution of the \hi\ gas mass content in galaxies from $z\approx 6$ to the present. In Section~\ref{sec:disc} we apply the [\cii]-to-\hi\ conversion to determine the comoving mass density of \hi\ in galaxies as a function of redshift $\rho_{\rm HI}(z)$ based on the luminosity functions of [\cii]. We provide a summary of our results in Section~\ref{sec:conc} and reflect on the future outlook of studying the \hi\ content of high-redshift galaxies.      

\begin{figure*}[t!]
\centering
\includegraphics[width=0.9\textwidth]{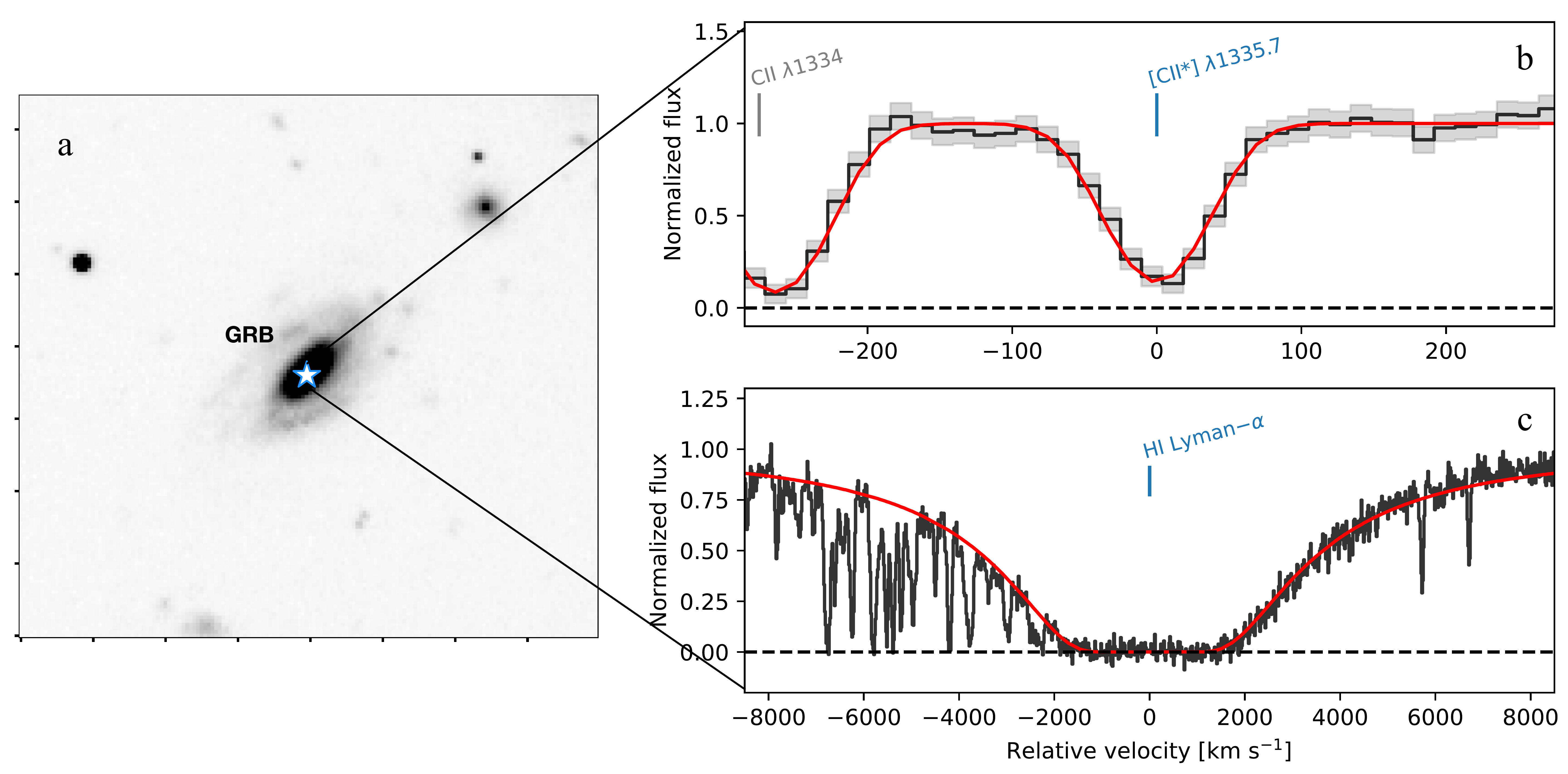}
\caption{Illustration of GRBs as probes of the interstellar medium in their hosts. Panel $a$ shows the typical explosion sites of GRBs, close to the center or brightest regions of their host galaxies. Panels $b$ and $c$ show examples of absorption features from [\cii*]\,$\lambda 1335.7$ and \hi\ Lyman-$\alpha$ imprinted on the optical afterglow spectrum of one of the sample GRBs, GRB\,090926A. Any velocity components of [\cii*]\,$\lambda 1335.7$, representing individual gas complexes along the line of sight, are typically resolved in wavelength space. The high abundance of \hi\ visualises as a single broad absorption trough. We use a sample of 15 GRB afterglows with good quality spectra to determine the [\cii*] and \hi\ content.}
\label{fig:galaxyillu}
\end{figure*}


\section{High-redshift [\cii]-to-\hi\ conversion} \label{sec:data}

\subsection{ESO-VLT/X-shooter observations}

The bright optical afterglows of GRBs can serve as powerful background sources to probe the various gas-phase constituents of the ISM in their host galaxies \citep{Jakobsson04,Fynbo06,Prochaska07}. Since long-duration GRBs are linked to the death of massive stars \citep{Woosley06}, they are expected to trace star-formation through cosmic time \citep{Robertson12,Tanvir12,Greiner15,Perley16}. Most GRB afterglow spectra are characterized by a broad absorption trough from the \hi\ Lyman-$\alpha$ transition \citep{Vreeswijk04,Jakobsson06,Fynbo09}, in line with being located behind a significant column of neutral gas in star-forming regions. Here we utilize the accurate abundances of \hi\ and [\cii*]\,$\lambda1335.7$, and the gas-phase metallicities, that can be derived in the line-of-sight through the ISM of GRB host galaxies to determine an empirical, metallicity-dependent [\cii]-to-\hi\ conversion factor in high-redshift galaxies. This is illustrated in Figure~\ref{fig:galaxyillu}.

This work is primarily based on the X-shooter GRB (XS-GRB) afterglow legacy survey \citep{Selsing19}. This sample includes a set of carefully-selected bursts detected by the Neil Gehrels {\it Swift} Observatory \citep{Gehrels04}, during a $\approx 10$ year period. The sample criteria are constructed such that the observed sample provides an unbiased representation of the underlying population of {\it Swift}-detected bursts, while simultaneously optimizing the observability. We consider all GRB afterglows suitable to measure the column densities of the \hi\ Lyman-$\alpha$ and [\cii*]\,$\lambda1335.7$ transitions, effectively requiring that the bursts occurred in galaxies at $z\gtrsim 2$. This corresponds to the wavelength range above which Lyman-$\alpha$ is observable from the ground due to the atmospheric cut-off at shorter wavelengths. The spectra are further required to have signal-to-noise ratios (S/N) in the wavelength region encompassing the two transitions of ${\rm S/N}\gtrsim 3$ to obtain reliable column density estimates. The final GRB afterglow sample considered in this work consists of the 15 bursts presented in Table~\ref{tab:grbtab}. While biases in how GRBs trace star-forming galaxies might be inherent at low redshifts \citep{Perley16,Palmerio19}, they have been demonstrated to be reliable tracers of star formation \citep{Robertson12,Greiner15} and sample the underlying star-forming galaxy population \citep{Perley16} at $z>2$.


\begin{table*}[t!]
\centering
\caption{GRB line-of-sight \hi\ and [\cii*]\,$\lambda 1335.7$ column densities and metal abundances.}
\begin{tabular*}{0.9\textwidth}{l @{\extracolsep{\fill}} ccccc}  
\hline
GRB & $z_{\rm GRB}$ & $\log N_{\rm HI}$ & $\log N_{\rm [CII^*]}$ & $\log(Z/Z_{\odot})$ & $\log (\beta_{\rm CII})$\\
 & & (cm$^{-2}$) & (cm$^{-2}$) & & ($M_{\odot}/L_{\odot}$) \\
\hline
090809A   &  2.7373 & $21.48\pm 0.07$ & $15.30\pm 0.39$ & $-0.46\pm 0.15$ & $2.25\pm 0.40$  \\
090926A   &  2.1069 & $21.58\pm 0.01$ & $14.67\pm 0.04$ & $-1.72\pm 0.05$ & $2.98\pm 0.04$ \\
100219A   &  4.6676 & $21.28\pm 0.02$ & $14.87\pm 0.13$ & $-1.16\pm 0.11$ & $2.48\pm 0.13$ \\
111008A   &  4.9910 & $22.39\pm 0.01$ & $16.01\pm 0.23$ & $-1.79\pm 0.10$ & $2.45\pm 0.23$ \\
120327A   &  2.8143 & $22.07\pm 0.01$ & $15.61\pm 0.47$ & $-1.34\pm 0.02$ & $2.53\pm 0.47$ \\
120716A   &  2.4874 & $21.73\pm 0.03$ & $15.81\pm 0.31$ & $-0.57\pm 0.08$ & $1.99\pm 0.31$ \\
120815A   &  2.3582 & $22.09\pm 0.01$ & $15.02\pm 0.08$ & $-1.23\pm 0.03$ & $3.14\pm 0.08$ \\
120909A   &  3.9290 & $21.82\pm 0.02$ & $16.01\pm 0.26$ & $-0.29\pm 0.10$ & $1.88\pm 0.26$ \\
121024A   &  2.3005 & $21.78\pm 0.02$ & $14.99\pm 0.24$ & $-0.68\pm 0.07$ & $2.86\pm 0.24$ \\
140311A   &  4.9550 & $22.30\pm 0.02$ & $15.02\pm 0.54$ & $-2.00\pm 0.11$ & $3.35\pm 0.54$ \\
151021A   &  2.3297 & $22.14\pm 0.03$ & $15.87\pm 0.50$ & $-0.97\pm 0.07$ & $2.34\pm 0.50$ \\
151027B   &  4.0650 & $20.54\pm 0.07$ & $14.62\pm 0.06$ & $-0.59\pm 0.27$ & $1.99\pm 0.09$ \\
160203A   &  3.5187 & $21.74\pm 0.02$ & $15.10\pm 0.31$ & $-0.92\pm 0.04$ & $2.71\pm 0.31$ \\
161023A   &  2.7100 & $20.95\pm 0.01$ & $14.79\pm 0.01$ & $-1.05\pm 0.04$ & $2.23\pm 0.01$ \\
170202A   &  3.6456 & $21.53\pm 0.04$ & $15.19\pm 0.24$ & $-1.02\pm 0.13$ & $2.41\pm 0.24$ \\
\hline
\end{tabular*}
\tablecomments{The \hi\ column densities and dust-corrected gas-phase metallicities are from \citet{Bolmer19}. [\cii*]\,$\lambda 1335.7$ column densities are derived in this work. $\beta_{\rm CII}$ denotes the [\cii]-to-\hi\ conversion factor per unit column derived for each system.}
\label{tab:grbtab}
\end{table*}


The GRBs in this sample probe galaxies spanning redshifts from $z=2.11$ (GRB\,090926A) to $z=4.99$ (GRB\,111008A). For each system, we adopt the \hi\ column densities and dust-corrected metallicities derived by \citet{Bolmer19}. The latter is computed from a range of metal species X tracing the neutral interstellar medium. The relative abundances for each element are derived as [X/H] $\equiv \log(N(\rm X)/N(\rm H))-\log(X/H)_{\odot}$, with ${\rm (X/H)_{\odot}}$ representing solar abundances \citep{Asplund09}. Since a fraction of the metals in the ISM are depleted by condensation onto interstellar dust grains, these are taken into account by correcting for the observed depletion level [X/Y], which is also correlated with the metallicity of the galaxy \citep{DeCia16}. This provides an estimate of the total, dust-corrected metallicity [X/H] + [X/Y] = [M/H], hence denoted as $\log (Z/Z_\odot)$. The GRB sample probes galaxies with \hi\ column densities in the line-of-sight in excess of $N_{\rm HI} = 2\times 10^{20}$\,cm$^{-2}$, classifying them as damped Lyman-$\alpha$ absorbers \citep{Wolfe05}, and relative abundances ranging from $\log (Z/Z_{\odot}) = -2.0$ to $\log (Z/Z_\odot) = -0.3$ (i.e. gas-phase metallicities of 1-50\% solar).

\subsection{Absorption-line fitting}

In this work we determine the column densities of the fine-structure [\cii*]\,$\lambda1335.7$ transition for each system. Since \cii\ traces the same neutral interstellar gas as several other, typically weaker transitions of other singly-ionized elements such as Fe\,{\sc ii}, Zn\,{\sc ii}, Si\,{\sc ii}, and S\,{\sc ii}, these are used to guide the velocity structure, number of components, and broadening parameter when fitting the [\cii*]\,$\lambda1335.7$ line complex. The absorption line profiles are modelled using \texttt{VoigtFit} \citep{Krogager18}, which simultaneously fits a set of Voigt profiles to the observed absorption features and provides the redshift $z_{\rm abs}$, column density $N$ and broadening parameter $b$ as output. For this procedure, the intrinsic line profiles are first deconvolved by the measured spectral resolution of each afterglow spectrum. For systems where multiple velocity components are detected, representing individual gas complexes along the line-of-sight, the sum of the individual column densities is reported. This is to be consistent with the procedure used to measure the \hi\ abundances and gas-phase metallicities. The resulting column densities $N_{\rm [CII^*]}$ are listed in Table~\ref{tab:grbtab}. We emphasize here that since the velocity structure of [\cii*]\,$\lambda1335.7$ in individual systems matches well the elements tracing the neutral gas-phase, and not the typical single-components observed for H$_2$ or [\ci] \citep{Bolmer19,Heintz19}, is further support of the physical connection between [\cii] and \hi.

\subsection{[\cii]-to-\hi\ calibration} \label{ssec:grbciitohi}

The line transition [\cii*]\,$\lambda1335.7$, typically observed in quasar DLA \citep{Wolfe03,Neeleman15} and GRB \citep{Fynbo09,Christensen11} absorption spectra, arises from the $^2P_{3/2}$ level in the ground-state of ionized carbon (C$^+$). Determining the column density of this particular feature ($N_{\rm [CII^*]}$) thus provides a measure of the C$^+$ in the $J=3/2$ state along the line-of-sight. The population in this state gives rise to the [\cii]-158\,$\mu$m transition ($^2P_{3/2}-^2P_{1/2}$). This allows us to determine the line's ``column'' luminosity $L^{\rm c}_{\rm [CII]}$, i.e. the luminosity per unit area, based on the spontaneous decay rate, as $L^{\rm c}_{\rm [CII]} = h\nu_{\rm ul}A_{\rm ul}N_{\rm [CII^*]}$. For [\cii]-158\,$\mu$m, $\nu_{\rm ul} = 1900.537$\,GHz and $A_{\rm ul} = 2.4\times 10^{-6}$\,s$^{-1}$. Similarly, the line-of-sight \hi\ column mass density can be determined, $M^{\rm c}_{\rm HI} = m_{\rm HI}N_{\rm HI}$, where $m_{\rm HI}$ is the mass of a single hydrogen atom and $N_{\rm HI}$ is total \hi\ column number density. This yields an expression for the line-of-sight [\cii]-to-\hi\ conversion factor in terms of the measured column densities $\beta_{\rm [CII]} \equiv M_{\rm HI}/L_{\rm [CII]} = M^{\rm c}_{\rm HI}/L^{\rm c}_{\rm [CII]} = m_{\rm HI}N_{\rm HI}/(h\nu_{\rm ul}A_{\rm ul}N_{\rm [CII^*]})$. Assuming that the ratio of the derived column densities for each sightline is representative of the mean of the relative total population, $N_{\rm HI}/N_{\rm [CII^*]} = \Sigma_{\rm HI}/\Sigma_{\rm [CII^*]}$, the $\beta_{\rm [CII]}$ calibration derived per unit column is thus equal to the global [\cii]-to-\hi\ conversion factor. This scaling is derived for each GRB in the sample and converted to solar units, $M_\odot/L_\odot$. Moreover, the absorption lines in GRB afterglow spectra also allow for very accurate measurements of the gas-phase metallicity to be made. We note that the column density of [\cii*]\,$\lambda1335.7$ has previously been used to estimate the total [\cii] luminosity of a DLA galaxy \citep{Simcoe20}, though based on strong assumptions about the geometry of the system. 

The conversion from the column density ratio to typical units of $M_{\odot}/L_{\odot}$ can be expressed as a constant scaling $M_{\rm HI}/L_{\rm [CII]} = N_{\rm HI}/N_{\rm [CII^*]} \times 1.165\times 10^{-4}\,M_{\odot}/L_{\odot}$. We wish to emphasize here, that a similar approach to determine the [\ci]- and CO-to-H$_2$ ratio in absorption-selected galaxies was demonstrated to be able to reproduce equivalent conversion factors observed in emission and expected from hydrodynamical simulations \citep{HeintzWatson20}, lending further credibility to this method. Moreover, this approach does not necessarily require the [\cii]-158\,$\mu$m emission to physically trace the \hi\ gas, it simply provides a measure of the total \hi\ gas and [\cii*] abundance in the line-of-sight, the latter also being partially produced in molecular clouds and photo-dissociation regions (PDRs).
Finally, our approach does not determine the global properties for the GRB-selected galaxies, only the relative mass to luminosity ratio of \hi\ and [\cii] respectively. This scaling, however, can be applied to infer the global \hi\ gas content of a given galaxy based on the total integrated luminosity of the [\cii] emission line. This is the methodology used throughout this work.

\begin{figure}[t!]
\centering
\includegraphics[width=8.7cm]{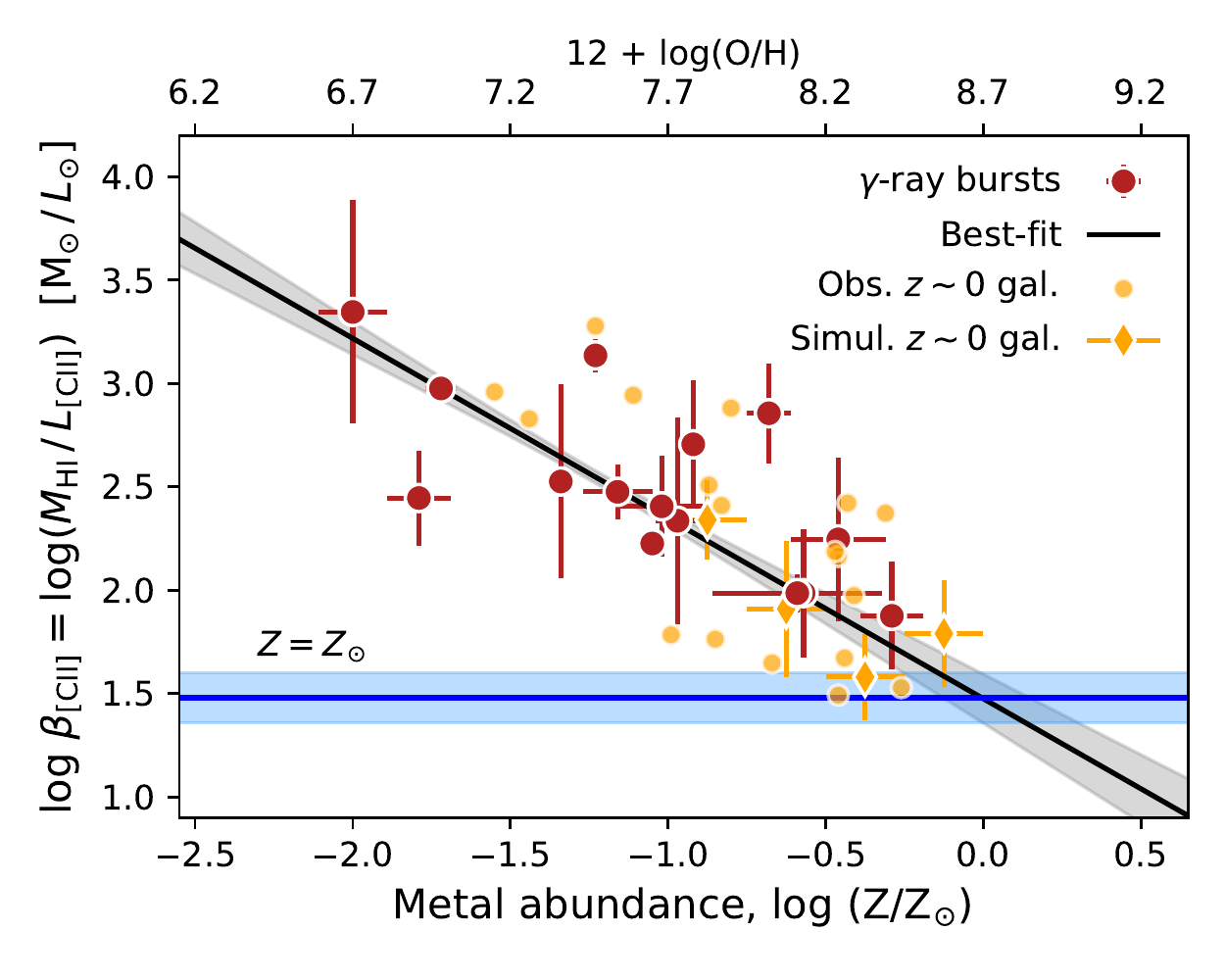}
\caption{Absorption-derived metallicity evolution of the \hi\ gas mass to [\cii] line luminosity. The red symbols show the measured relative column densities of \hi\ Lyman-$\alpha$ and [\cii*]\,$\lambda1335.7$ in the line-of-sight, converted to units of $M_{\odot}/L_{\odot}$, for each GRB. The gas-phase metal abundances have been measured from absorption lines and are corrected for dust. The black solid line and grey-shaded region represent the best-fit linear relation and associated uncertainty. The blue solid line and shaded region mark the extrapolated [\cii]-to-\hi\ conversion factor $\beta_{\rm [CII]}$ at solar metallicities, $\beta_{\rm [CII]}=30^{+10}_{-7}\,M_{\odot}/L_{\odot}$.
For comparison, the gray dots show the relative [\cii] luminosity to \hi\ gas mass inferred from 21-cm observations of galaxies at $z\sim 0$ and the dark-gray diamond symbols show the average $M_{\rm HI}/L_{\rm [CII]}$ ratios in metallicity bins of 0.25 dex for a set of simulated galaxies at $z\sim 0$ (see text).}
\label{fig:ciitohi}
\end{figure}

Figure~\ref{fig:ciitohi} shows the measured [\cii]-to-\hi\ relative abundances as a function of gas-phase metallicity, including the best-fit scaling relation between the two. The conversion factor, $\beta_{\rm [CII]} = M_{\rm HI}/L_{\rm [CII]}$, is found to be linearly dependent on the metallicity, described by the following relation
\begin{equation}
\begin{split}
    \log M_{\rm HI} = (-0.87\pm 0.09) \times \log(Z/Z_{\odot}) + \\ (1.48\pm 0.12) + \log L_{\rm [CII]} 
    \label{eq:ciitohi}
\end{split}
\end{equation}
where $Z/Z_{\odot}$ is the relative solar abundance (with $12+\log({\rm O/H})_{\odot} = 8.69$ for $\log (Z/Z_{\odot}) = 0$) \citep{Asplund09}, and $M_{\rm HI}$ and $L_{\rm [CII]}$ are in units of $M_{\odot}$ and $L_{\odot}$, respectively. 

The observed scatter is likely dominated by variations in the physical properties of the ISM (such as density and gas pressure) and the intensity of the ultraviolet (UV) background field \citep{Popping19}. For comparison, we overplot the sample of galaxies at $z\sim 0$ from the {\it Herschel} Dwarf Galaxy Survey \citep{Madden13}, for which [\cii] luminosities and \hi\ gas masses from direct 21-cm observations have been inferred \citep[see][and references therein]{RemyRuyer14,Cormier15}. We here only consider the main-sequence galaxies from this sample (within 0.5 dex), as parametrized by \citet{Speagle14}, to be consistent with the main analysis described in Sect.~\ref{sec:res}. Additionally, we include the average $M_{\rm HI}/L_{\rm [CII]}$ ratios of a set of $z\sim 0$ simulated galaxies are shown in metallicity bins of 0.25 dex based on the simulations of \citet{Olsen21}. These theoretical expectations and the $z\sim 0$ galaxy sample are observed to match well with the empirical linear relation of $\beta_{\rm [CII]}$ as a function of metallicity observed for the GRB sample at $z>2$, indicating a strong universal connection between the two.

The same analysis and results could in principle also be reproduced for DLAs originating in foreground galaxies toward bright background quasars. However, since quasar-selected DLAs are typically observed at high impact parameters \citep{Peroux11,Krogager12,Christensen14,Rahmani16,Krogager17}, they mostly probe the extended \hi\ envelope rather than the gas directly responsible for star formation in their associated galaxies \citep{Neeleman19}. At these high impact parameters, the [\cii] emission may be shock-heated, intergalactic gas \citep{Appleton13} rather than excited by star formation. As a consequence, the observed [\cii*] contribution is likely to be much lower per unit \hi\ gas mass. Indeed, our preliminary analysis of quasar-DLAs show on average a factor of $\approx 10$ lower [\cii] ``column'' luminosity pr unit \hi\ ``column'' mass. DLA galaxy counterparts also show suppressed star formation compared to the ``main-sequence'' of star-forming galaxies at equivalent redshifts \citep{Rhodin18}, indicating that quasar-DLA gas is not typically representative of the gas in star-forming galaxies.

\section{Results} \label{sec:res}

For the [\cii]-to-\hi\ conversion factor to be applicable to high redshift galaxies, it is important that representative galaxy samples with high-quality auxiliary data are surveyed. For this work, we use recent sample compilations of [\cii]-emitting galaxies at two distinct epochs: $z\sim 2$ and $z\sim 4-6$. These epochs are mainly defined by the redshifted frequency of the [\cii]-158\,$\mu$m transition observable by the available ALMA bands.
Starbursts and quasar host galaxies are excluded in this analysis, such that only main-sequence star-forming galaxies are considered (see specific references below). Auxiliary data for each galaxy are further required; at minimum measurements of the star-formation rate (SFR) and $M_\star$. These measurements are used to infer the gas-phase metallicity of each galaxy from the recent calibration of the fundamental metallicity relation (FMR) \citep{Curti20}. This is parametrized as
\begin{equation}
    Z(M_\star,{\rm SFR}) = Z_0 - \gamma / \beta \log (1 + (M_\star/M_0({\rm SFR}))^{-\beta})
\end{equation}
where $M_0({\rm SFR}) = 10^{m_0}\times {\rm SFR}^{m_1}$, where we adopt their best-fitting parameters considering the total SFR: $Z_0 = 8.779$, $m_0 = 10.11$, $m_1 = 0.56$, $\gamma = 0.31$, and $\beta = 2.1$.
Gas-phase metallicities for each galaxy are required to accurately determine the \hi\ gas mass (Eq.~\ref{eq:ciitohi}). While the FMR is not well-established at $z>3$, the average metallicities inferred for the high-redshift samples described below reproduces the expected decrease of $\Delta \log(Z/Z_\odot)=-0.1$ per unit redshift observed for star-forming galaxies up to $z\approx 3.5$ \citep{Sanders20}, and potentially beyond \citep{Cucchiara15,Faisst16,Jones20fmr}.

\subsection{Galaxy sample compilation}

For the main analysis, a survey for [\cii] emission in main-sequence galaxies at $z\sim 2$ \citep{Zanella18} is included to represent [\cii]-emitting star-forming galaxies at this epoch. At higher redshifts, $z\sim 4-6$, we make use of the recent ALPINE survey \citep{LeFevre20,Bethermin20,Faisst20,Fujimoto20} which includes measurements of the [\cii] line emission and kinematics, and physical galaxy properties based on extensive UV to sub-mm data. Supplementing this survey, we include an additional sample of galaxies at $z\sim 5-6$ from \citet{Capak15}. The full high-redshift sample is comprised of main-sequence star-forming galaxies with stellar masses in the range $M_{\star} = 10^{9.5}-10^{11}\,M_{\odot}$, SFRs of $\approx 5$ to 600\,$M_{\odot}\,{\rm yr}^{-1}$, and gas-phase metallicities $12+\log{\rm (O/H)} = 8.12$ to 8.77 ($Z=0.3-1.2\,Z_\odot$). 

Due to the accessibility of the 21\,cm line transition as a tracer of \hi\ in the local Universe, these observations are utilized here to represent the \hi\ gas content of galaxies at $z=0$. As the reference sample at $z\sim 0$, we adopt the combined catalog of galaxies from The \hi\ Nearby Galaxy Survey \citep[THINGS;][]{Walter08} and the HERA CO-Line Extragalactic Survey \citep[HERACLES;][]{Leroy08}. This sample includes galaxies with stellar masses, SFRs, and atomic and molecular hydrogen gas mass estimates. Importantly, it includes galaxies with a mass range similarly distributed as those in the high-$z$ sample compilation. For consistency, the FMR is used to compute the gas-phase metallicity for each system, yielding metal abundances in the range $12+\log{\rm (O/H)} = 8.6-8.8$ (i.e. $80\%-125$\% solar abundance). In our analysis, we also include the recent estimates of the average \hi\ gas content at $z\approx 1-1.3$ \citep{Chowdhury20,Chowdhury21}, based on stacking of the \hi\ 21-cm signal of a large ensemble of star-forming galaxies. This work provides measurements of $M_{\rm HI}$ and $\rho_{\rm HI}$ intermediate to the galaxy samples at $z\sim 0$ and $z\gtrsim 2$ examined here.

\begin{figure}[t!]
\centering
\includegraphics[width=8.7cm]{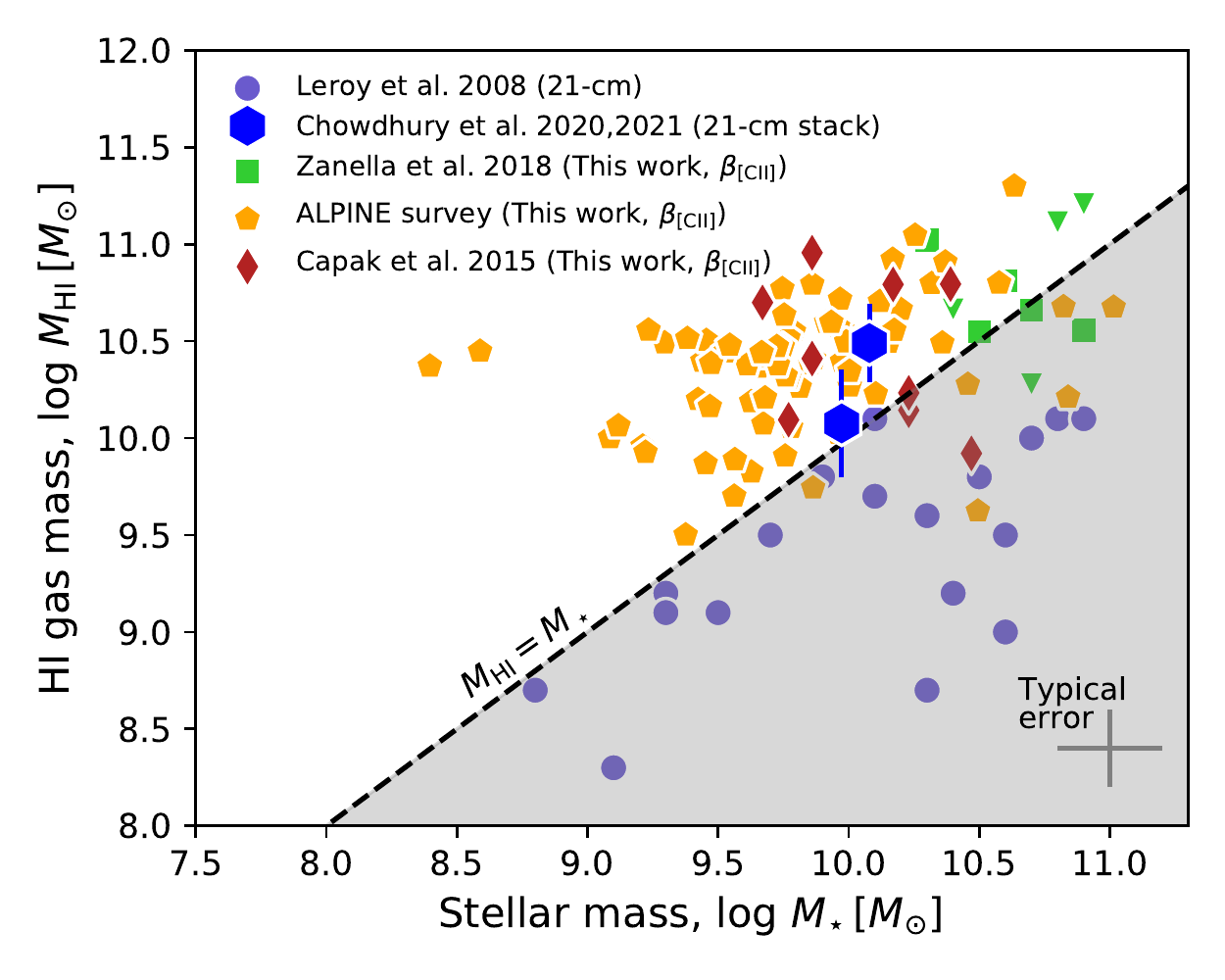}
\caption{\hi\ gas mass and stellar mass distributions. The compiled galaxy samples are shown as a function of redshift, with distinct symbols and colors. The new measurements of the \hi\ gas masses at $z>2$ were derived using the [\cii]-to-\hi\ conversion factor from Figure~\ref{fig:ciitohi}, based on high-redshift [\cii]-emitting galaxy samples: green: \citet{Zanella18}, yellow: The ALPINE survey \citep{LeFevre20,Bethermin20,Faisst20}, and red: \citet{Capak15}. Lower redshift measurements come from 21-cm observations: purple: \citet[][$z\approx 0$]{Leroy08}, blue: \citet[][$z\approx 1$]{Chowdhury20,Chowdhury21}. The dashed line represent $M_{\rm HI} = M_\star$, clearly separating galaxies with \hi\ gas masses in excess (majority at $z\gtrsim 1$) or in deficit ($z\sim 0$) of the stellar mass. }
\label{fig:mhimstar}
\end{figure}

For the compiled set of [\cii]-emitting galaxies at $z=2-6$ we infer \hi\ gas masses in the range $M_{\rm HI} = 3\times 10^{9} - 2\times 10^{11}\,M_\odot$. At redshifts $\approx 1$ and above, all galaxies are observed to have \hi\ gas masses in excess of their stellar mass $M_\star$, contrasting the low \hi\ content in nearby galaxies (see Figure~\ref{fig:mhimstar}). The high-redshift \hi\ mass estimates are consistent with the inferred excess of the total ISM mass for galaxies in a similar stellar mass range \citep[typically by a factor of $2-3$ over the stellar mass; e.g.][]{Scoville17}, the first indication that \hi\ dominates the total ISM mass of galaxies at $z\gtrsim 2$. 

\subsection{Evolution of the \hi\ gas fraction with redshift}

\begin{figure}[t!]
\centering
\includegraphics[width=8.7cm]{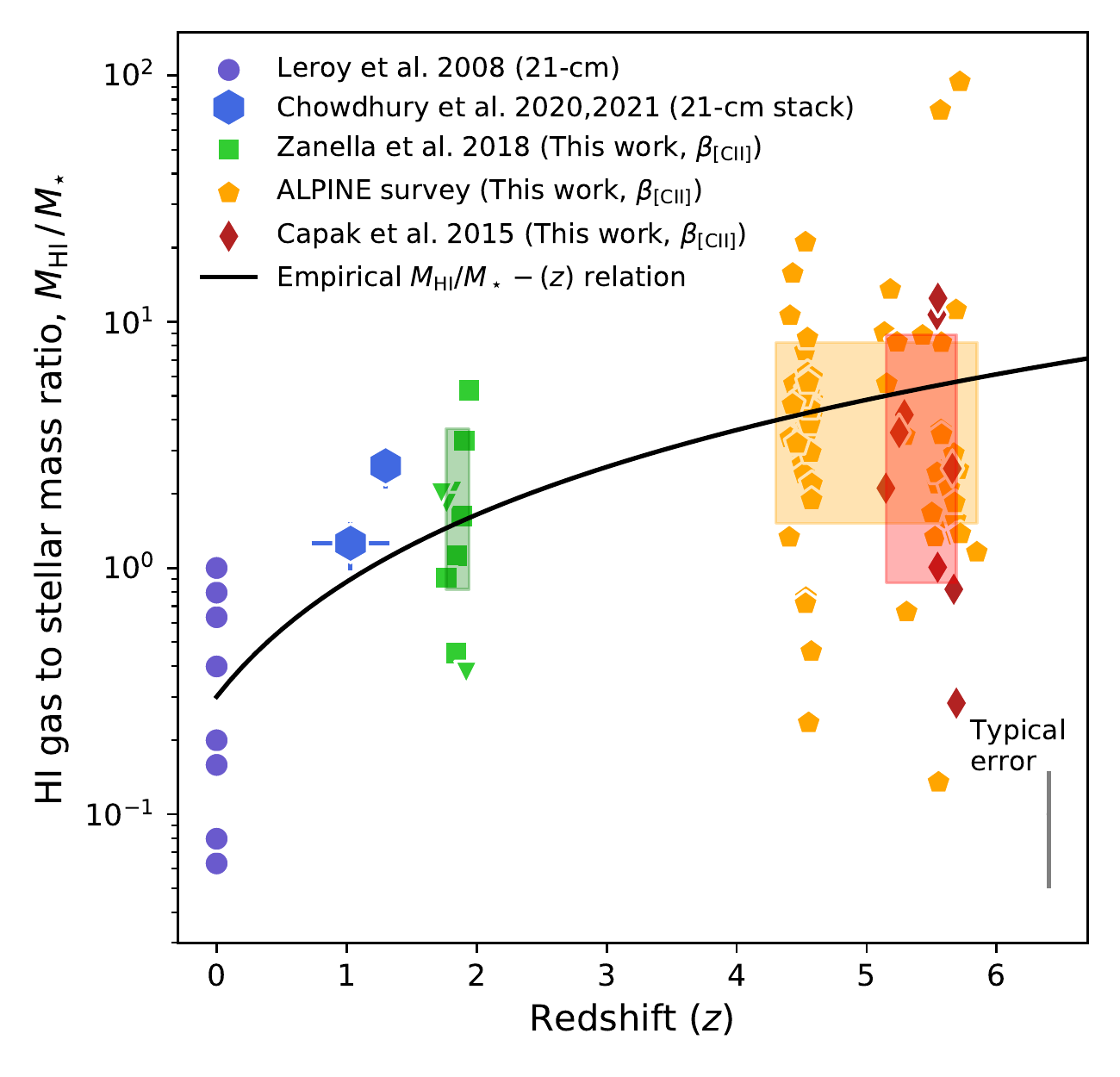}
\caption{Redshift evolution of the \hi\ gas mass of star-forming galaxies. Symbol notation follow Figure~\ref{fig:mhimstar}. Colored boxes mark the 16th to 84th confidence interval of $M_{\rm HI}/M_{\star}$ and span the full redshift distribution for each high-redshift sample. The black curve shows the expected evolution of $M_{\rm HI}/M_{\star}(z)$ based on independent empirical scaling relations (see text).}
\label{fig:mhimstarz}
\end{figure}

With estimates of the \hi\ content in galaxies from $z\approx 6$ to the present, we first quantify the redshift evolution of the \hi\ gas fraction, $M_{\rm HI}/M_\star$. The full compiled data set is shown in Figure~\ref{fig:mhimstarz}, where the \hi\ gas fraction is observed to increase from $M_{\rm HI}/M_\star = 0.37^{+0.42}_{-0.29}$ ($1\sigma$ confidence interval of sample distribution) at $z=0$ to $M_{\rm HI}/M_\star = 6.79^{+1.45}_{-5.27}$ at $z\sim 4-6$. Coupling the $z\sim 0-1$ data based on 21-cm observations of \hi\ with the new measurements based on the [\cii]-to-\hi\ conversion factor we observe a steady increase of $M_{\rm HI}/M_\star$ as a function of redshift. 
Equivalently, the fraction of \hi, $f_{\rm HI} = M_{\rm HI}/(M_{\rm HI}+M_\star)$, is observed to constitute $\approx 70\%$ at $z\sim 4-6$, decreasing to $f_{\rm HI} \approx 25\%$ at $z=0$. 


We confirm the determined redshift evolution of the \hi\ gas fraction by constructing a simple model using literature scaling relations. Here, we subtract the measured average trend of $M_{\rm H_2}/M_\star$ (i.e. $M_{\rm H_2}/M_\star \propto (1+z)^{2.3}$, \citealt{Carilli13,Tacconi18}) from the measured evolution of the total ISM mass (i.e. $M_{\rm gas} = M_{\rm H_2} + M_{\rm HI}$), prescribed as $M_{\rm gas}/M_\star \propto (1+z)^{1.84}$ \citep{Scoville17}. This yields a function of the form
\begin{equation}
    \frac{M_{\rm HI}}{M_\star} = 0.4\times (1+z)^{1.84} \left[1-\frac{(1+z)^{0.46}}{4}\right]
\end{equation}
based on the observed ratio $M_{\rm H_2} = 1/3 \, M_{\rm HI}$ for the reference sample at $z=0$, shown in Figure~\ref{fig:mhimstarz}. This expression can be approximated by $M_{\rm HI}/M_\star = 0.3 \times (1+z)^{1.55}$. The strong correspondence between the inferred trend in this simple model and our measurements is striking. We note that this observed trend indicates that \hi\ is more abundant than H$_2$ at all redshifts, when compared to the evolution of $M_{\rm H_2} / M_{\rm HI}$ with redshift \citep[e.g.,][]{Geach11,Carilli13,Tacconi18}.

Moreover, while not directly measurable from the available data, we can still provide an approximate estimate of the fraction of \hi\ to the total baryonic mass (i.e. $M_{\rm bar,tot} = M_{\rm HI} + M_{\rm H_2} + M_\star$) from these relations. Starting from $M_{\rm H_2} = 1/3 \, M_{\rm HI}$ at $z=0$ and accounting for the predicted redshift evolution of $M_{\rm H_2}/M_{\rm HI} \propto (1+z)^{-0.34}$ \citep{Morselli21}, we estimate $M_{\rm HI} / M_{\rm bar,tot} \approx 60\%$ at $z\sim 4-6$, which decreases to $\approx 20\%$ at $z=0$. We obtain similar results if we instead adopt the [\cii]-to-H$_2$ scaling from \citet{Zanella18} to additionally compute $M_{\rm H_2}$. Based on this and the results above, \hi\ is thus observed to dominate the baryonic matter content of star-forming galaxies at $z\gtrsim 2$.

Finally, while [\cii] is now being increasingly detected in emission, well into the epoch of reionization at $z\gtrsim 7$ \citep[e.g.,][]{Knudsen17,Bradac17,Smit18,Hashimoto19,Fujimoto19}, we refrain from including these systems in the main analysis due to the lack of well-defined samples surveying [\cii] at these redshifts. The [\cii]-to-\hi\ conversion factor derived is still applicable for these distant galaxies though. For a few of the individual [\cii]-emitters detected at $z\sim 7$ \citep{Knudsen17,Smit18}, we infer \hi\ masses exceeding the stellar mass by a factor of $\approx 10$, consistent with the measured redshift evolution of the $M_{\rm HI}/M_{\star}$ ratio.

\subsection{Metallicity-dependence of the \hi\ gas fraction}

\begin{figure}[t!]
\centering
\includegraphics[width=8.7cm]{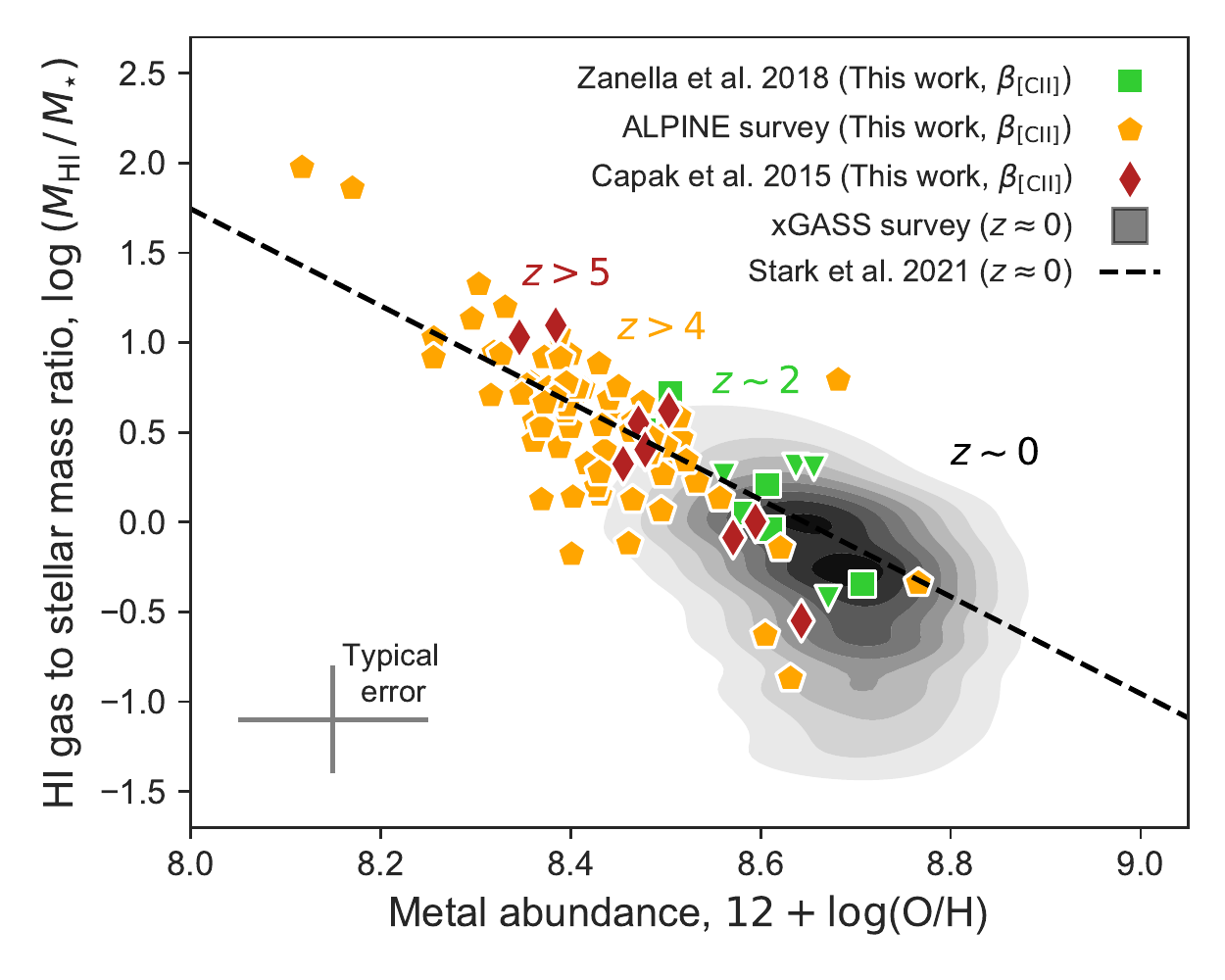}
\caption{Metallicity evolution of the \hi\ gas fraction, $M_{\rm HI}/M_\star$. Symbol notation follow Figure~\ref{fig:mhimstar}. Metal abundances are determined for each high-redshift galaxy based on the FMR \citep{Curti20}. The xGASS catalog of galaxies at $z\sim 0$ with direct \hi\ 21-cm observations \citep{Catinella18} is shown as the grey-shaded contours for comparison, together with the empirical relation derived for the \hi-MaNGA sample galaxies \citep{Stark21} (black dashed line). Both nearby and distant galaxies, at $z=0-6$, are observed to follow this fundamental relation.}
\label{fig:mhimstarmet}
\end{figure}

In addition to examining the evolution of the \hi\ gas fraction with redshift, here we quantify whether the high-redshift measurements follow the observed anti-correlation with gas-phase metallicity observed in the local Universe \citep{Hughes13,Brown18,Stark21}. At fixed redshifts, this is equivalently represented as a decrease in the \hi\ fraction with increasing stellar mass \citep{Catinella13,Brown15}.
In Figure~\ref{fig:mhimstarmet} we show the gas fraction $M_{\rm HI}/M_\star$ as a function of metal abundance $12+\log$(O/H) for each galaxy in our compiled high-$z$ sample. For comparison, we overplot the xGASS catalog of galaxies at $z\sim 0$ with direct \hi\ 21-cm observations \citep{Catinella18}, together with the empirical relation derived for the \hi-MaNGA sample galaxies \citep{Stark21}. The combined sample of galaxies, spanning redshifts from $z=0-6$, is observed to closely follow this local, empirical relation. 

This establishes the relation between the gas fraction $M_{\rm HI}/M_\star$ and $12+\log$(O/H) as fundamental. Moreover, it further substantiates the empirical [\cii]-to-\hi\ conversion factor derived here and validates the inferred \hi\ gas masses of the high-redshift galaxy samples.

\subsection{\hi\ depletion times}

With the inferred \hi\ gas masses, we can also estimate the \hi\ depletion times, $t_{\rm dep,HI} = M_{\rm HI}/{\rm SFR}$, of each galaxy, which provides the time-scale on which \hi\ would be exhausted by star-formation. This estimate is independent of whether the neutral gas has to first be converted to H$_2$ or can form stars directly from the available neutral gas. At $z\sim 0$, star-forming galaxies typically have \hi\ depletion times $t_{\rm dep,HI} \approx 5$\,Gyr, significantly longer than the H$_2$ depletion times observed for the same galaxies ($t_{\rm dep,H_2} \lesssim 1$\,Gyr, \citealt{Saintonge17}). At $z\approx 1$, $t_{\rm dep,HI}$ decreases to an average about $1.5$\,Gyr \citep{Chowdhury20}. 

\begin{figure}[t!]
\centering
\includegraphics[width=8.9cm]{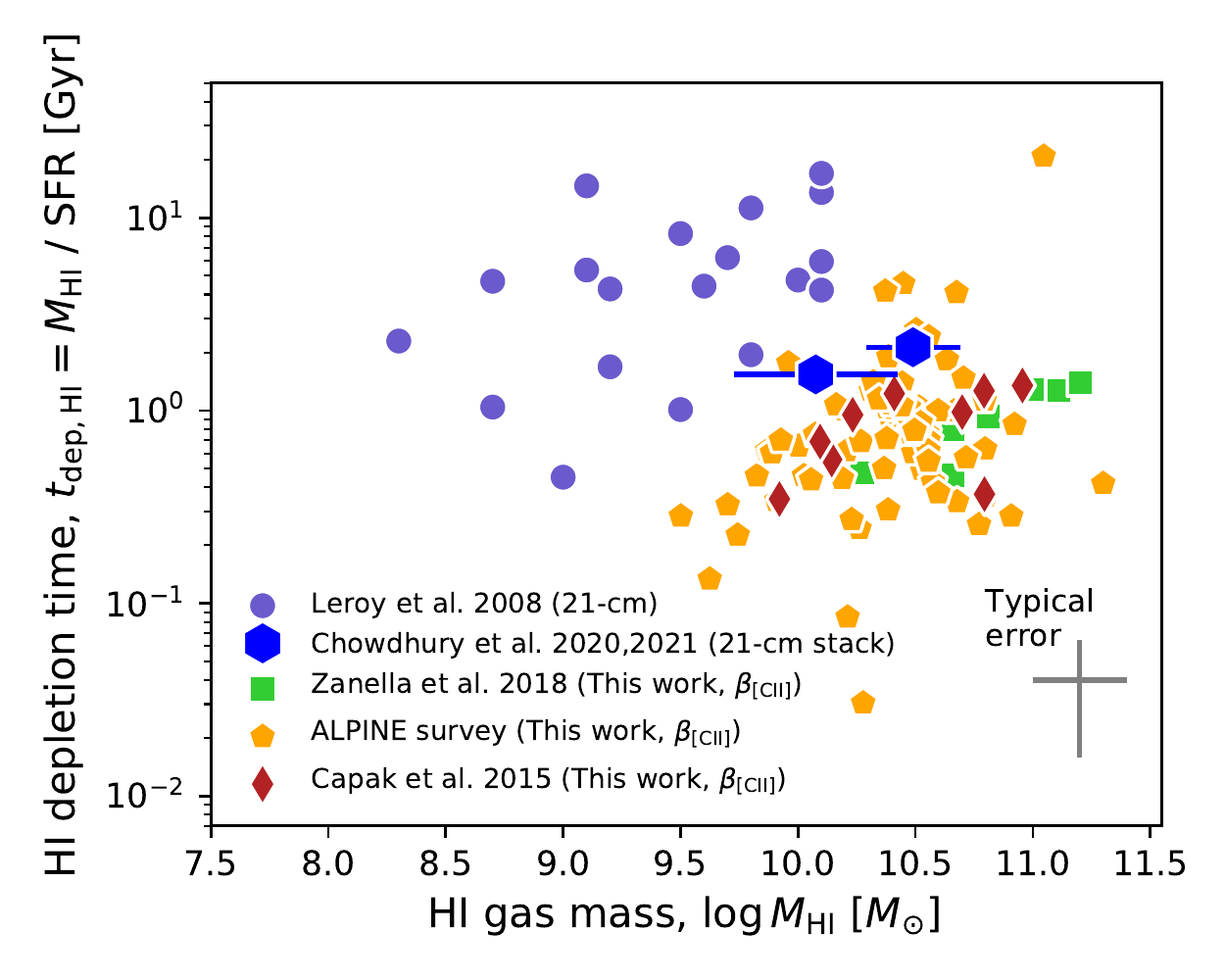}
\caption{\hi\ depletion time of star-forming galaxies. Symbol notation follow Figure~\ref{fig:mhimstar}. Galaxies at $z\sim 0$ have available neutral gas reservoirs to maintain star formation for $\approx 1-10$\,Gyr. The majority of galaxies at higher redshifts have \hi\ depletion times $\lesssim 2$\,Gyr.}
\label{fig:hidepl}
\end{figure}

In Figure~\ref{fig:hidepl}, we explore the \hi\ depletion timescales for the high-redshift galaxy samples, including the lower redshift comparison samples with direct \hi\ 21-cm measurements. As already hinted by the $z\approx 1-1.3$ measurements from \citet{Chowdhury20,Chowdhury21}, we confirm a continued decrease of $t_{\rm dep,HI}$ at $z\gtrsim 2$. At these redshifts, the majority of galaxies show \hi\ depletion times $\lesssim 2$\,Gyr. This indicates that at least for the first 4 billion years of cosmic time, the \hi\ gas reservoirs in galaxies has to be constantly replenished to fuel the high star-formation activity. When this infall of neutral gas is discontinued, the available gas will be exhausted due to the short \hi\ depletion timescales at $z\gtrsim 2$, thereby quenching the on-going star formation as a result. 

\begin{figure}[t!]
\centering
\includegraphics[width=8.7cm]{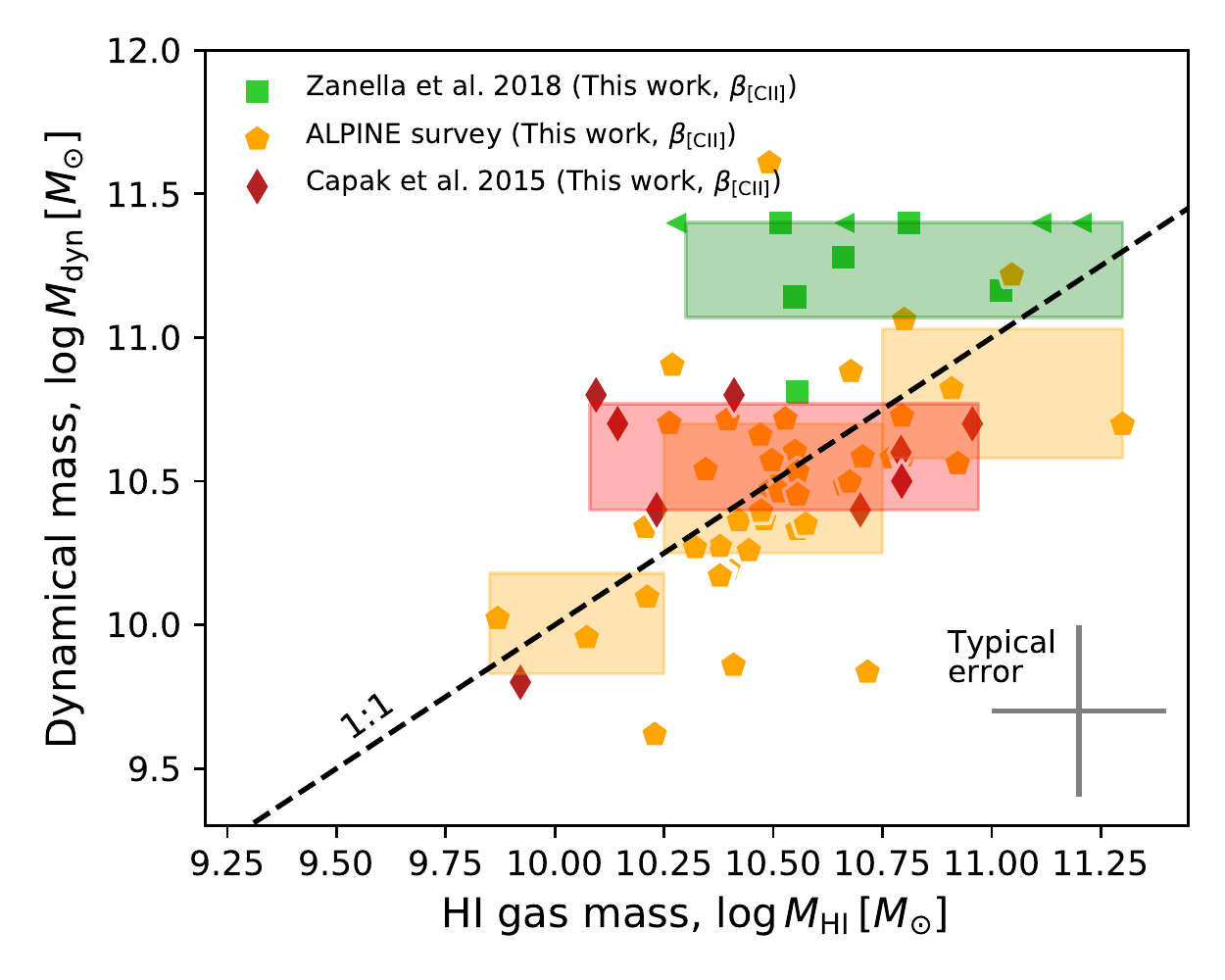}
\caption{Comparison of the dynamical mass and \hi\ gas mass estimated from the [\cii] line. Boxes mark vertically the 16th-84th percentile on the dynamical mass distribution of each sub-sample and horizontally the \hi\ gas mass interval. Colors of boxes correspond to the samples they represent. The general good consistency between $M_{\rm dyn}$ and $M_{\rm HI}$ of the galaxies at $z\gtrsim 4$ indicates that the \hi\ gas contributes dominantly to the overall dynamic mass within the [\cii]-emitting region. Galaxies at $z\sim 2$ show larger dynamical masses than the \hi\ component, possibly due to larger contributions from the molecular gas and stellar component at this epoch.}
\label{fig:mhimdyn}
\end{figure}

\subsection{Comparison to the dynamical mass}

To further explore the contribution of $M_{\rm HI}$ to the total ISM mass budget in high-redshift galaxies, we here compare our measurements to the dynamical mass $M_{\rm dyn}$ estimates of each galaxy. Assuming the [\cii] emission originates from a rotation-dominated disk we infer $M_{\rm dyn}$ within the [\cii]-emitting region determined from the individual line widths. We determine this as $M_{\rm dyn} = 1.16\times 10^{5}\,v_{\rm cir}^2\,D$, with $D$ the disk diameter and $v_{\rm cir}=0.75\,{\rm FWHM}_{\rm [CII]}/\sin(i)$ \citep{Wang13}. For the cases where the inclination angle $i$ of the galaxy is not available, we assume $i=55^\circ$ \citep{Wang13,Willott15}. The dynamical masses are shown in Figure~\ref{fig:mhimdyn}, in comparison to the \hi\ gas mass. For the $z\sim 2$ galaxy sample, $M_{\rm dyn}$ is on average $\approx 3$ times higher than $M_{\rm HI}$, consistent with a larger contribution from the molecular gas and stellar mass components in galaxies at these redshifts. The galaxies at $z\gtrsim 4$, however, on average show dynamical masses consistent with the inferred \hi\ mass. 

We note that \citet{DessaugesZavadsky20} found consistent results between the molecular gas masses inferred for a subset of the ALPINE sample and the total dynamical mass after accounting for the stellar mass contribution. However, these estimates are based on the $\alpha_{\rm [CII]}$ calibration derived for solar-metallicity galaxies at $z\approx 2$ by \citet{Zanella18}, which have not been verified for galaxies with lower gas-phase metallicities. It is clear though that both the atomic and molecular gas potentially both present a large mass fraction.   

These considerations and the results above are yet more evidence indicating that \hi\ dominates the baryonic matter content of galaxies at this epoch. Moreover, these results lend further credibility to the accuracy of the [\cii]-to-\hi\ calibration, demonstrating that it can recover sensible values over a large mass range.


\section{The cosmic \hi\ gas mass density} \label{sec:disc}

The cosmic density of neutral atomic gas associated with galaxies ($\rho_{\rm HI}$) is an important parameter in understanding the baryon cycle of galaxies. In the local Universe, $\rho_{\rm HI}$ can be derived from 21-cm \hi\ observations of individual galaxies \citep{Zwaan05,Hoppmann15,Jones18HI}. This method can be extended to $z\approx 1$ by considering an ensemble of galaxies, i.e.\ via stacking \citep{Chowdhury20}. At higher redshifts, $z\gtrsim 2$, $\rho_{\rm HI}$ has been measured from quasar absorption spectroscopy of damped Lyman-$\alpha$ (DLAs) systems \citep{Peroux03,Prochaska09,Noterdaeme12,Neeleman16}. DLAs only probe random sightlines through their associated galaxy counterparts, however, and are increasingly difficult to detect at $z\gtrsim 4$ due to the blanketing effect of the Lyman-$\alpha$ forest. In most cases, DLA sightlines also only recover the extended \hi\ envelope of their associated galaxies rather than the gas responsible for star formation \citep[][see also Sect.~\ref{ssec:grbciitohi}]{Neeleman19}. 

\begin{figure*}[t!]
\centering
\includegraphics[width=16cm]{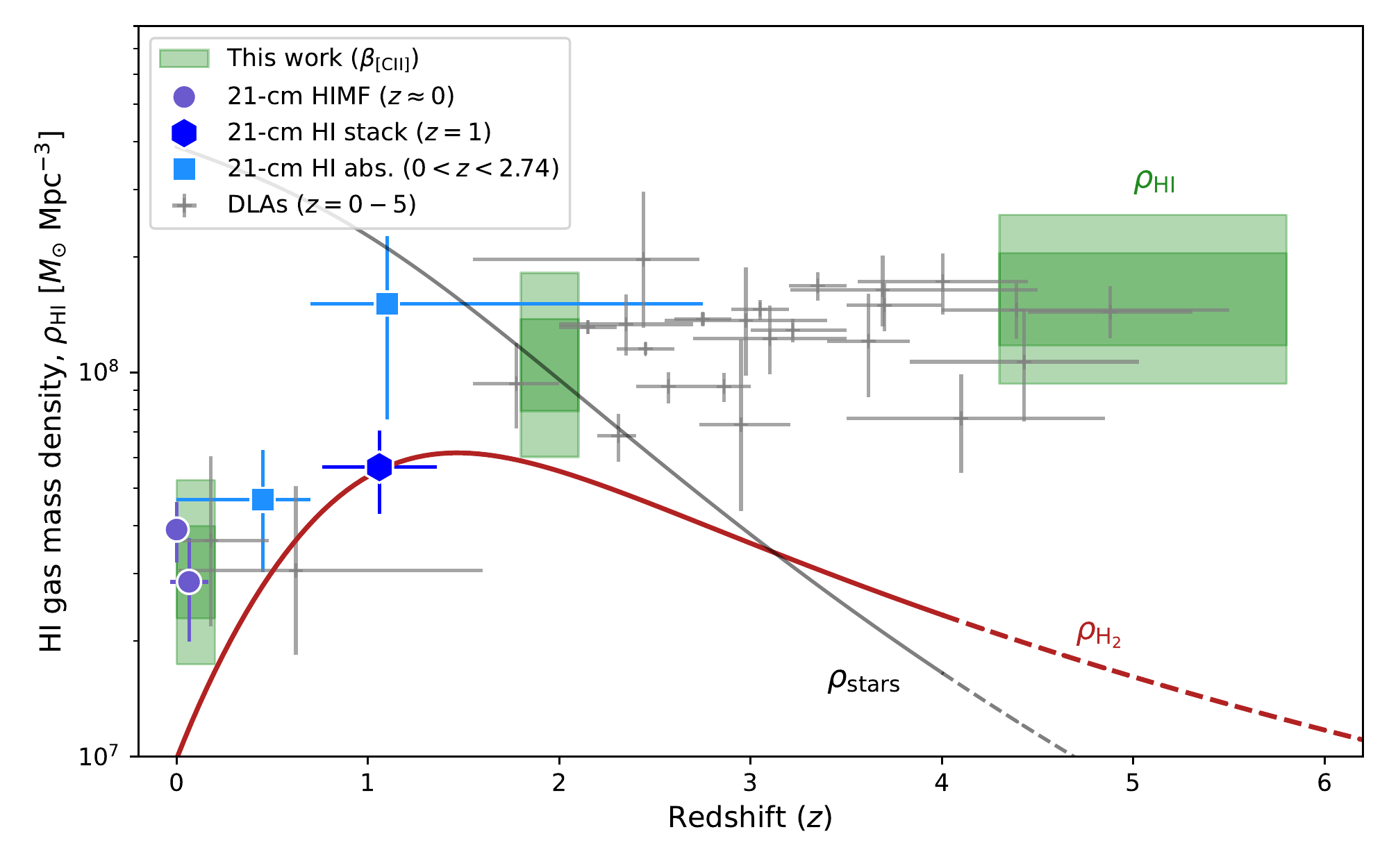}
\caption{Redshift evolution of the cosmic \hi\ gas mass density in galaxies $\rho_{\rm HI}$. Measurements from this work are based on the estimated [\cii] luminosity densities at each sample redshift, converted to $\rho_{\rm HI}$ (see text). The vertical dark and light green regions encompass the $68\%$ and $95\%$ confidence levels, respectively, of the $\rho_{\rm HI}$ measurements and horizontally indicate redshift intervals. The gray dots at $z\sim 0$ show measurements of $\rho_{\rm HI}$ based on \hi\ mass functions from recent 21-cm observations \citep{Hoppmann15,Jones18HI} and the large hexagon shows the average measurement of an ensemble of galaxies at $z\approx 1$ \citep{Chowdhury20}. Blue squares show the results from a survey of intervening 21-cm \hi\ absorption systems covering $0z<z<0.69$ and $0.69<z<2.74$ \citep{Grasha20}. Grey crosses represent DLA measurements \citep{Walter20}, corrected by $20\%$ to account for the contribution from sub-DLAs and Lyman-limit systems. Our measurements are consistent with these other recent estimates and indicate (at $\approx 3\sigma$ significance) a decrease of $\rho_{\rm HI}$ in galaxies by a factor of $\approx 5$ from $z\sim 4-6$ to the present. For comparison, the evolution of the cosmic density of stars\cite{Madau14} and molecular gas in galaxies are shown as well, extrapolated beyond $z>4$ from the analysis of \citet{Walter20}.}
\label{fig:rhohi}
\end{figure*}

The [\cii]-to-\hi\ conversion factor derived here allows us to measure $\rho_{\rm HI}$ directly from the star-forming gas in galaxies at $z\approx 2$ and beyond. To do this, we derive the [\cii] luminosity density, $\mathcal{L}_{\rm [CII]}$, by integrating over the entire luminosity function $\mathcal{L}_{\rm [CII]} = \int^{\infty}_{L_{\rm [CII]}} L_{\rm [CII]} \phi(L_{\rm [CII]}) dL_{\rm [CII]}$. We separate the analysis into three sample redshifts; $z=4-6$, $z=2$, and $z=0$. This is to account for the evolution in the luminosity function of [\cii] and the overall increase in metal abundance of typical galaxies across cosmic time. A similar approach is commonly used to determine the H$_2$ gas mass density in galaxies $\rho_{\rm H_2}(z)$ based on the relevant CO luminosity function and a CO-to-H$_2$ conversion factor \citep{Walter14,Decarli16,Decarli19,Decarli20,Riechers19}. 
At the highest redshifts, $z=4-6$, we adopt the [\cii] luminosity function from the ALPINE survey \citep{Yan20}. To infer the equivalent cosmic \hi\ gas mass density in galaxies, the average metallicity $12+\log{\rm (O/H)} = 8.4$ (i.e. $\log(Z/Z_{\odot}) = -0.3$) observed for the sample galaxies at the same redshift is used to determine the relevant conversion factor of $\beta_{\rm [CII]} = 55^{+18}_{-13}\,M_{\odot}/L_{\odot}$. This yields $\rho_{\rm HI} = 1.6^{+0.5}_{-0.4}\times 10^{8}\,M_{\odot}\,{\rm Mpc^{-3}}$ at $z\sim 4-6$. Since the  infrared luminosity function, which is connected to [\cii], shows little evolution from $z=6$ to $z\approx 2$ \citep{Gruppioni20,Loiacono21}, we adopt the same [\cii] luminosity density as above to estimate $\rho_{\rm HI}$ at $z=2$. However, since the average metal abundance of galaxies increases with decreasing redshift, a higher average metallicity is required to determine the appropriate $\beta_{\rm [CII]}$ calibration. Based on the average gas-phase metallicity of $12+\log{\rm (O/H)} = 8.6$ (i.e. $\log(Z/Z_{\odot}) = -0.1$) for the subset of galaxies at $z\approx 2$, we derive a conversion factor of $\beta_{\rm [CII]} = 37^{+12}_{-9}\,M_{\odot}/L_{\odot}$. This results in $\rho_{\rm HI} = 1.0^{+0.3}_{-0.3}\times 10^{8}\,M_{\odot}\,{\rm Mpc^{-3}}$ at $z=2$, consistent (within $1\sigma$) with the results at $z\sim 4-6$. Due to a significant evolution of the [\cii] luminosity function from $z=6$ to the present, the more appropriate function from \citet{Hemmati17} is adopted at $z\sim 0$. Integrating this yields a luminosity density $\mathcal{L}_{\rm [CII]} = 1.0\times 10^6\,L_{\odot}{\rm Mpc}^{-3}$. At solar abundances, the average inferred for the subset of galaxies at $z\approx 0$, the [\cii]-to-\hi\ conversion factor is $\beta_{\rm [CII]} = 30^{+10}_{-7}$. This translates to a local measurement of $\rho_{\rm HI} = 3.0^{+1.0}_{-0.7} \times 10^{7}\,M_{\odot}\,{\rm Mpc^{-3}}$.

Figure~\ref{fig:rhohi} shows the measured redshift evolution of $\rho_{\rm HI}(z)$. For comparison, we only include the most recent estimates of $\rho_{\rm HI}$ derived from \hi\ 21-cm observations of local galaxies \citep{Hoppmann15,Jones18HI}. We note that previous estimates from 21-cm observations seem to systematically recover higher 
values of $\rho_{\rm HI}$ at $z\sim 0$ \citep[e.g.][]{Lah07,Braun12}. The most recent 21-cm measurements, however, are also in best agreement with low-redshift DLA measurements \citep{Neeleman16,Shull17}, so we only consider these for the main analysis. In the figure we also include the inferred average value of $\rho_{\rm HI} = (5.7\pm 1.4) \times 10^{7}\,M_{\odot}\,{\rm Mpc^{-3}}$ in star-forming galaxies at $z\approx 1$ from \citet{Chowdhury20}. We also show the results from \citet{Grasha20}, surveying 21-cm \hi\ absorption systems at $z=0-2.74$ in addition to the DLA sample compiled by \cite{Walter20}, including $\rho_{\rm HI}$ measurements covering $z\sim 0-5$ from previous works \citep{Peroux03,Prochaska09,Guimaraes09,Crighton15}, but here corrected by 20\% to account for the contribution to the cosmic \hi\ budget from sub-DLAs and Lyman-limit systems \citep{OMeara07,Noterdaeme12,Zafar13,Berg19}.

We observe that the \hi\ density in galaxies shows a negligible decrease in the first $\approx 4$ Gyr after the Big Bang, from $z\approx 6$ to $z=2$. At this point, coincident with the peak of cosmic star-formation rate density, the cosmic \hi\ mass density declines substantially by a factor of $\approx 5$ to the present. This decline is a factor of $\approx 2$ larger than previously inferred \citep[e.g.][]{Walter20,Peroux20}. 
In contrast, the stellar mass density in galaxies $\rho_{\rm stars}$ increase by a factor of $\approx 50$ during the same cosmic time span. Intriguingly, at $z=0$, $\rho_{\rm stars}$ exceeds the sum of the baryonic densities at $z=6$, suggesting continued accretion of gas onto galaxies from outside the regions we measure with [\cii]. The significant decrease in $\rho_{\rm HI}$ (and $\rho_{\rm H_2}$), however, indicates that the neutral atomic gas in galaxies begins to be exhausted at this point. The paucity of continuous \hi\ gas infall and accretion, required to replenish the available gas reservoirs, is thus not sufficient to fuel and maintain star formation following the peak of the cosmic star-formation rate density. 
These results further support a decreasing fraction of $\rho_{\rm H_2} / (\rho_{\rm H_2} + \rho_{\rm HI})$ beyond $z\approx 2$ \citep{Walter20}, reaching 10\% at $z\sim 4-6$. This is contrary to recent results suggesting an increase in the molecular to total ISM gas fraction \citep{Sommovigo21}.  

Finally, we can also make a simple prediction of the redshift evolution of $\rho_{\rm HI}$ based on these new data points. A simple function of the form $\rho_{\rm HI}(z) = 3\times 10^{7}(1+z)^{0.85}\,M_{\odot}\,{\rm Mpc^{-3}}$ (or equivalently $\Omega_{\rm HI}(z) = 2\times 10^{-4}(1+z)^{0.85}$) is found to represent the data well. Notably, this predicts a significant decrease in $\rho_{\rm HI}(z)$, with a value at $z=0$ that is approximately a factor of two lower than previously inferred \citep{Walter20,Peroux20}.

\section{Summary and Outlook} \label{sec:conc}

This work provides the first direct calibration of the conversion between the [\cii] line luminosity $L_{\rm [CII]}$ and the \hi\ gas mass $M_{\rm HI}$ for high-redshift galaxies, here denoted $\beta_{\rm [CII]} \equiv M_{\rm HI}/L_{\rm [CII]}$. The atomic gas-phase in the ISM of galaxies at these epochs is otherwise undetectable due to the weakness of the typically employed tracer, the fine-structure 21-cm \hi\ transition, which only allow observations of individual galaxies out to $z\approx 0.4$ \citep{Fernandez16} or $z\approx 1$ by combining the signals from thousands of galaxies \citep{Chowdhury20} with current radio observatories.  

We applied this calibration to existing samples of [\cii]-emitting galaxies at $z\approx 2-6$ to quantify the evolution of the \hi\ gas content in galaxies through cosmic time. We found that the \hi\ gas fraction, $M_{\rm HI}/M_\star$, increases continuously as a function of redshift, with $M_{\rm HI}$ exceeding $M_\star$ at $z\approx 1$. The \hi\ gas fraction was found to similarly be dependent on the gas-phase metallicity, $12+\log({\rm O/H})$, following a universal anti-correlation from $z=0$ to $z=6$. Comparing the inferred \hi\ gas masses to the measured SFRs of each galaxy, we uncovered that the majority of galaxies at $z\gtrsim 2$ have \hi\ depletion timescales, $t_{\rm dep,HI} = M_{\rm HI}/{\rm SFR}$, less than $\approx 2$\,Gyr, substantially shorter than observed in galaxies at $z\approx 0$. This we argued indicates that the atomic gas reservoirs is required to be constantly replenished at $z\gtrsim 2$ to fuel the high star-formation activity. 
We further compared the inferred \hi\ gas masses to the total dynamical masses $M_{\rm dyn}$ derived from the individual line widths within the [\cii]-emitting region. At $z\sim 4-6$, $M_{\rm HI}$ was found to be consistent with $M_{\rm dyn}$ on average, suggesting that \hi\ dominates the total ISM mass budget at this epoch, whereas a larger contribution to $M_{\rm dyn}$ from other components, likely the molecular gas, was evident in galaxies at $z\approx 2$.  

We further used the $\beta_{\rm [CII]}$ calibration to determine the comoving mass density of \hi\ in galaxies as a function of redshift $\rho_{\rm HI}(z)$. This was based on measurements of the [\cii] luminosity density $\mathcal{L}_{\rm [CII]}$ at three distinct epochs, $z=0$, $z=2$, and $z=4-6$, using previously constrained luminosity functions, which can be converted into $\rho_{\rm HI}$. We measured $\rho_{\rm HI} = 1.6^{+0.5}_{-0.4}\times 10^{8}\,M_{\odot}$\,Mpc$^{-3}$ at $z\sim 4-6$, a factor of $\approx 5$ larger than at $z=0$. These measurements were consistent with previous estimates using DLAs. We emphasize, however, that the $\beta_{\rm [CII]}$ calibration has the potential to infer $\rho_{\rm HI}$ far beyond current estimates, well into the epoch of reionization at $z\approx 7$. 

As already demonstrated by surveys targeting CO and \ci\ as tracers of the molecular gas content, the use of the [\cii]-to-\hi\ conversion factor derived here is similarly important to advance our understanding of the atomic gas content in high-redshift galaxies. Even with next generation radio observatories such as the Square Kilometer Array (SKA), the \hi\ 21-cm transition might only be detected for individual galaxies out to $z\approx 1.7$ \citep{Blyth15}, further motivating the development and use of alternative tracers. We encourage follow-up observations and simulations to further substantiate the [\cii]-to-\hi\ conversion derived here.

\section*{Acknowledgements}

We would like to thank the referee for constructive feedback on the analysis and presentation of the results in this work. We thank Johan P. U. Fynbo and J. Xavier Prochaska for insightful discussions during an early stage of this project. We thank Karen P. Olsen for early access to the simulated galaxy data set. We thank Marcel Neeleman for providing us the DLA sample compilation. K.E.H. acknowledges support by a Postdoctoral Fellowship Grant (217690--051) from The Icelandic Research Fund. D.W. is supported by Independent Research Fund Denmark grant DFF--7014-00017. The Cosmic Dawn Center is funded by the Danish National Research Foundation under grant No.~140. D.N. acknowledges support from the U.S. NSF via grant AAG--1909153. \\

\section*{Data availability statement} 

The ESO-VLT/X-shooter GRB afterglow spectra are all publically available through the ESO archive in the form of phase 3 material. The ALMA-ALPINE catalog is available at \url{https://cesam.lam.fr/a2c2s/data_release.php}. The open-source code {\tt VoigtFit} used to model the absorption-line features is available at: \url{https://github.com/jkrogager/VoigtFit}. Source codes for the figures and tables presented in this manuscript are available from the corresponding author upon reasonable request.\\


\bibliography{ref}
\bibliographystyle{aasjournal}


\end{document}